\documentclass[preprint,prc,showpacs,preprintnumbers,amsmath,amssymb,floatfix]{revtex4}
\usepackage{amsmath}

\usepackage{graphicx}
\usepackage{dcolumn}
\usepackage{bm}
\usepackage{ulem} 
\usepackage[usenames]{color}
\usepackage{epstopdf}
\usepackage{epsfig}
\usepackage{float}
\usepackage{subfigure}
\usepackage{multirow}
\usepackage[version=3]{mhchem}

\newcommand{\nc}{\newcommand}       
\nc{\vc}[1] {\mbox{\boldmath $#1$}} 
\nc{\del}       {\partial}              
\nc{\bra}       {\langle}               
\nc{\ket}       {\rangle}               
\nc{\bras}[1]   {\langle #1|}           
\nc{\kets}[1]   {|#1\rangle}            
\nc{\mapleft}[1]{           
 \smash{\mathop{\,          %
  \hbox to 1.5cm{\rightarrowfill}\, }\limits_{#1}}}
\nc{\beq}     {\begin{eqnarray}} \nc{\eeq}    {\end{eqnarray}}
\nc{\nn}      {\\\nonumber} \nc{\vs}      {\vspace{-0.275cm}}
\nc{\fra}    {\frac{1}{2}}
\nc{\mb}        {\mathbf}

\begin{document}

\preprint{}

\title{The quark mean field model with pion and gluon corrections}

\author{Xueyong Xing}
\affiliation{School of Physics, Nankai University, Tianjin 300071,  China}
\author{Jinniu Hu}
\email{hujinniu@nankai.edu.cn}
\affiliation{School of Physics, Nankai University, Tianjin 300071,  China}
\author{Hong Shen}
\affiliation{School of Physics, Nankai University, Tianjin 300071,  China}

\date{\today}
\begin{abstract}
The properties of nuclear matter and finite nuclei are studied within the quark mean field (QMF) model by taking the effects of pion and gluon into account at the quark level.  The nucleon is described as the combination of three constituent quarks confined by a harmonic oscillator potential. To satisfy the spirit of QCD theory, the contributions of pion and gluon on the nucleon structure are treated in second-order perturbation theory. For the nuclear many-body system, nucleons interact with each other by exchanging mesons between quarks. With different constituent quark mass, $m_q$, we determine three parameter sets about the coupling constants between mesons and quarks, named as QMF-NK1, QMF-NK2, and QMF-NK3 by fitting the ground-state properties of several closed-shell nuclei. It is found that all of the three parameter sets can give satisfactory description on properties of nuclear matter and finite nuclei, meanwhile they can also predict the larger neutron star mass around $2.3M_\odot$ without the hyperon degrees of freedom.
\end{abstract}

\pacs{21.10.Dr,  21.60.Jz,  21.80.+a}

\keywords{Quark mean field, Pion, Gluon}

\maketitle

\section{Introduction}
The nucleon, as the element of nuclear physics, is composed of a more microscopic particle, which is named as quark in the standard model. Quarks are confined in the nucleon and interact with each other through exchanging the gluon in the basic theory of strong interaction, Quantum Chromodynamics (QCD). Such confinement will increase hugely with the distance between two quarks. Therefore, it is impossible to observe the free quarks in universe except at some extreme conditions such as high density and high temperature.

At a nuclear energy level, the interaction between quarks cannot be treated with perturbation theory, whose coupling strength is not small anymore. Until now, we still could not describe the nucleon structure as a few-body problem based on the degrees of freedom of quarks and gluons directly with QCD theory like $ab~initio$ calculation.  With the development of computer technology, the mass spectrum of baryons and mesons have been calculated with lattice QCD theory as a few-quarks system, which can treat the quark confinement very well with numerical methods. The mass spectrum of mesons and baryons was reproduced very well compared to experiment observations~\cite{jansen08}.

However, it is a big challenge to study the nuclear many-body system based on the degrees of freedom of quarks and gluons with QCD theory, which will help us understand the nuclear physics more fundamentally. To overcome such difficult problem, a lot of many-body methods were proposed, which can reasonably give the properties of finite nuclei from light mass region to heavy mass region very well, like Green Function Monto Carlo (GFMC) method~\cite{pieper01}, Shell model~\cite{cauries05}, Skyrme Hartree-Fock (SHF) theory~\cite{erler11}, relativistic mean field (RMF) theory~\cite{serot86} and so on. However, in most of these models, the nucleon is usually treated as a point particle, which does not consider its internal structure. This assumption is against the observation of present experiments about the nucleon structure. Furthermore, the medium modification of the nucleon structure function (EMC effect) cannot be achieved by treating the nucleon as a point particle. It is a challenge to develop a model for understanding the change of the nucleon structure in nuclei from quark degrees of freedom.

Therefore, Guichon proposed an exploratory model to study the nuclear many-body problem, where the quarks are confined in a bag, like the MIT bag model and interact with each other at different nucleons through exchanging of $\sigma$ and $\omega$ mesons~\cite{guichon88}. The mechanism of nuclear saturation properties can be discussed in terms of quark degrees of freedom. Later, Saito and Thomas \textit{et al.} extended such model to include more meson, like $\rho$ meson, called as quark-meson coupling (QMC) model~\cite{saito94,guichon96,saito97}. The QMC model can be regarded as an extension of the RMF model, where the scalar coupling constant, generated from the effective nucleon mass, is changed with the quark mass in nuclear medium. Actually, it corresponds to the nuclear EMC effect~\cite{aubert83}, which modifies the nucleon structure in nuclear medium. With the development of the QMC model persistently, it is not only applied on the study of nuclear structure, but also on the hadron physics, astrophysics and particle physics~\cite{panda97,panda99,panda03,panda04,panda12,saito07}.

The quark is treated as a current quark in QMC model, whose mass is very small, just several MeV. Accordingly, Toki \textit{et al.} used a constituent quark model for the nucleon instead of the MIT bag model~\cite{tok98}, named as quark mean field (QMF) model. The quarks are confined through some confinement potentials in the QMF model, whose constituent quark mass is around $300$ MeV. Shen \textit{et al.} promoted such picture with more precise parameters by fitting the properties of stable finite nuclei~\cite{shen00} and applied it on the study of hypernuclei and neutron stars~\cite{shen02,hu14a,hu14b}. To include the baryon octets, Wang \textit{et al.} introduced a chiral Lagrangian at hadron level in QMF model and studied the properties of strangeness nuclear matter~\cite{wang01,wang02,wang03,wang04a,wang04b,wang05}.

In the QMC and QMF models, two very important factors in QCD theory are neglected. One is the gluon, which propagates the interact of quarks. The other is the chiral symmetry, which is taken into account by pion and generates the quark mass from chiral symmetry limit. The gluon can interact with itself and the contribution of pion in mean field is zero. Many works firstly attempted to take the pion effect within QMF model through the Fock term~\cite{krein99,stone07,whittenbury14}. Recently, Nagai \textit{et al.} extended the QMC model to include the gluon and pion exchange effect by using the cloudy bag model (CBM)~\cite{nagai08}.  Furthermore, Barik \textit{et al.} calculated the contribution of quark-pion interaction and quark-gluon interaction by one-pion exchange and one-gluon exchange, respectively in lowest-order perturbation theory in 1980s~\cite{barik85,barik86}.

Recently, Barik \textit{et al.} developed their method from single baryon states to the infinite nuclear matter following the methodology of the QMC and QMF models, named as MQMC model, and discussed the properties of symmetric nuclear matter influenced by the quark mass with $\sigma$ and $\omega$ meson exchanges~\cite{barik13}. Later, they also included the $\rho$ meson to study the asymmetric nuclear mechanical instability and its dependence on the isospin asymmetry of the system~\cite{mishra15}. In their works, the free parameters, like the strengths of $\sigma$ and $\omega$ mesons are determined by the nuclear saturation properties. The non-linear terms of $\sigma$ and $\omega$ mesons were not included in the Lagrangian, which resulted in a large effective nucleon mass. Therefore, such parameter sets obtained by Barik \textit{et al.}  from nuclear matter could not describe the properties of finite nuclei very well.

Therefore, in this work, we would like to include the contribution of pion and gluon with the perturbation theory in QMF model and fix the free parameters from the ground state of stable finite nuclei. Then, we will apply such new parameters to study the properties of nuclear matter and neutron stars.

The paper is written as follows. In Sec. II, we briefly derive the contribution of pion and gluon on nucleon properties with perturbation theory and the formulas of QMF model on nuclear matter and finite nuclei. In Sec. III, the new parameter sets of QMF model with pion and gluon corrections will be determined. The properties of nuclear matter and finite nuclei with such new parameter sets will be shown. A summary is given in Sec. IV.

\section{Quark mean field model with pion and gluon corrections}
The analytical confinement potential for quarks cannot be obtained from QCD theory directly. Many phenomenological confinement potentials were proposed, where the polynomial forms were widely used. A harmonic oscillator potential, $U(r)$, is adopted in this work, with which the Dirac equation can be solved analytically~\cite{barik13},
\beq\label{1}
U(r)=\frac{1}{2}(1+\gamma^0)(ar^2+V_0),
\eeq
where, the scalar-vector form of the Dirac structure is chosen for the quark confinement potential. Here, we should emphasize that such potential is just used for the baryon state, not for meson state, since in QMF model, the degrees of freedom for quark and meson are equally treated. $a$ and $V_0$ are the potential parameters, which are determined by the free nucleon mass and radius. When the effect of nuclear medium is considered, the quark field $\psi_{q}(\vec{r})$ satisfies the following Dirac equation,
\beq\label{3}
&&[\gamma^{0}(\epsilon_{q}-g^q_\omega\omega-\tau_{3q}g^q_\rho\rho)-\vec{\gamma}\cdot\vec{p}\nn&&-(m_{q}
-g^q_\sigma\sigma)-U(r)]\psi_{q}(\vec{r})=0,
\eeq
where $\sigma$, $\omega$, and $\rho$ are the classical meson fields, which take the exchanging interaction between quarks. $~g^q_\sigma, ~g^q_\omega$, and $g^q_\rho$ are the coupling constants of $\sigma, ~\omega$, and $\rho$ mesons with quarks, respectively. $\tau_{3q}$ is the third component of isospin matrix,  and $m_q$ is the bare quark mass. Now we can define the following quantities for later convenience,
\beq\label{4}
\epsilon^{\prime}_q&=&\epsilon_q^*-V_0/2,\nn
m^{\prime}_q&=&m_q^*+V_0/2,
\eeq
where the effective single quark energy is given by, $\epsilon_q^*=\epsilon_{q}-g^q_\omega\omega-\tau_{3q}g^q_\rho\rho$ and the effective quark mass by $m_q^*=m_{q}-g^q_\sigma\sigma$~\cite{shen00}.  We also introduce $\lambda_q$ and $r_{0q}$ as
\beq\label{5}
\lambda_q&=&\epsilon^{\prime}_q+m^{\prime}_q,\nn
r_{0q}&=&(a\lambda_q)^{-\frac{1}{4}}.
\eeq
The nucleon mass in nuclear medium can be expressed as the binding energy of three quarks named as zeroth-order term, after solving the Dirac equation (\ref{3}), formally
\beq\label{6}
E_N^{*0}=\sum_q\epsilon^*_q.
\eeq

The quarks are simply confined in a two-body confinement potential. Three corrections will be taken into account in the zeroth-order nucleon mass in nuclear medium, including the center-of-mass correction $\epsilon_{\rm c.m.}$, the pion correction $\delta M_{N}^\pi$, and the gluon correction $(\Delta E_N)_g$. The pion correction is generated by the chiral symmetry of QCD theory and the gluon correction by the short-range  exchange interaction of quarks. The center-of-mass correction can be obtained ~\cite{barik13} from
\beq\label{7}
\epsilon_{\rm c.m.}=\bras{N}\mathcal{H}_{\rm c.m.}\kets{N},
\eeq
where $\mathcal{H}_{\rm c.m.}$ is the center of mass Hamiltonian density and $\kets{N}$ is the nucleon state. When the nucleon wave function is constructed by the quark wave functions, the center-of-mass correction comes out as
\beq\label{8}
\epsilon_{\rm c.m.}=\frac{77\epsilon^{\prime}_{q}+31m^{\prime}_{q}}{3(3\epsilon^{\prime}_{q}+m^{\prime}_{q})^{2}r^{2}_{0q}}.
\eeq
In order to restore the chiral symmetry in nucleon, an elementary pion field is introduced in the present model. Pion contribution is zero in first-order perturbation theory due to its pseudovector properties. Therefore, we should treat it with second-order perturbation theory. Then, the pionic self-energy correction on nucleon mass becomes,
\beq\label{9}
\delta M^{\pi}_{B}=-\sum\limits_k\sum\limits_{B^{\prime}}\frac{V^{\dag BB^{\prime}}_{j}V^{BB^{\prime}}_{j}}{w_{k}},
\eeq
where $\sum\limits_{k}\equiv\sum\limits_{j}\int d^{3}k/(2\pi)^{3}$, $w_{k}=(k^{2}+m_{\pi}^{2})^{1/2}$ is the pion energy, and $V^{BB^{\prime}}_{j}$
represents the baryon pion absorption vertex function in the point-pion approximation. Then it can be simplified as
\beq\label{10}
\delta M^{\pi}_{N}=-\frac{171}{25}I_{\pi}f_{NN\pi}^{2},
\eeq
where,
\beq\label{11}
I_{\pi}=\frac{1}{\pi m_{\pi}^{2}}\int_0^\infty dk\frac{k^{4}u^{2}(k)}{w_{k}^{2}},
\eeq
with the axial vector nucleon form factor,
\beq\label{12}
u(k)=\left[1-\frac{3}{2}\frac{k^{2}}{\lambda_{q}(5\epsilon^{\prime}_{q}+7m^{\prime}_{q})}\right]e^{-\frac{1}{4}r^{2}_{0q}k^2}
\eeq
and $f_{NN\pi}$ can be obtained from the  Goldberg-Triemann relation by using the axial-vector coupling-constant value $g_A$ in this model. The one-gluon exchange contribution to the mass is separated into two parts as
\beq\label{13}
(\Delta E_{B})_{g}=(\Delta E_{B})_{g}^{E}+(\Delta E_{B})_{g}^{M},
\eeq
where $(\Delta E_{B})_{g}^{E}$ is the color-electric contribution
\beq\label{14}
(\Delta E_{B})_{g}^{E}=\frac{1}{8\pi}\sum\limits_{i,j}\sum\limits_{a=1}^{8}\int\frac{d^{3}r_{i}d^{3}r_{j}}
{|\vec{r}_{i}-\vec{r}_{j}|}\bras{B}J^{0a}_{i}(\vec{r}_{i})J^{0a}_{j}(\vec{r}_{j})\kets{B},
\eeq
and $(\Delta E_{B})_{g}^{M}$ the color-magnetic contribution
\beq\label{15}
(\Delta E_{B})_{g}^{M}=-\frac{1}{8\pi}\sum\limits_{i,j}\sum\limits_{a=1}^{8}\int\frac{d^{3}r_{i}d^{3}r_{j}}
{|\vec{r}_{i}-\vec{r}_{j}|}\bras{B}\vec{J}^{a}_{i}(\vec{r}_{i})\cdot\vec{J}^{a}_{j}(\vec{r}_{j})\kets{B}.
\eeq
Here
\beq\label{16}
J^{\mu a}_{i}(x)=g_{c}\bar{\psi}_{q}(x)\gamma^{\mu}\lambda^{a}_{i}\psi_{q}(x)
\eeq
is the quark color current density, where $\lambda_i^a$ are the usual Gell-Mann SU(3) matrices and
$\alpha_{c}=g_{c}^{2}/4\pi$. Then Eqs. (\ref{14}) and (\ref{15}) can be written as
\beq\label{17}
(\Delta E_N)_g^E={\alpha_c}(b_{uu}I_{uu}^E+b_{us}I_{us}^E+b_{ss}I_{ss}^E),
\eeq
and
\beq\label{18}
(\Delta E_N)_g^M={\alpha_c}(a_{uu}I_{uu}^M+a_{us}I_{us}^M+a_{ss}I_{ss}^M),
\eeq
where $a_{ij}$ and $b_{ij}$ are the numerical coefficients depending on each baryon and the quantities $I_{ij}^{E}$ and $I_{ij}^{M}$ are given in the following equations
\beq\label{19}
I_{ij}^{E}&=&\frac{16}{3{\sqrt \pi}}\frac{1}{R_{ij}}\left[1-
\frac{\alpha_i+\alpha_j}{R_{ij}^2}+\frac{3\alpha_i\alpha_j}{R_{ij}^4}
\right],\nn
I_{ij}^{M}&=&\frac{256}{9{\sqrt \pi}}\frac{1}{R_{ij}^3}\frac{1}{(3\epsilon_i^{'}
+m_{i}^{'})}\frac{1}{(3\epsilon_j^{'}+m_{j}^{'})},
\eeq
with
\beq\label{20}
R_{ij}^{2}&=&3\left[\frac{1}{({\epsilon_i^{'}}^2-{m_i^{'}}^2)}+
\frac{1}{({\epsilon_j^{'}}^2-{m_j^{'}}^2)}\right],
\nn
\alpha_i&=&\frac{1}{ (\epsilon_i^{'}+m_i^{'})(3\epsilon_i^{'}+m_{i}^{'})}.
\eeq
Finally, taking above pion and gluon corrections, the mass of nucleon in nuclear medium becomes
\beq\label{21}
M_N^*=E_N^{*0}-\epsilon_{\rm c.m.}+\delta M_N^\pi+(\Delta E_N)^E_g+(\Delta E_N)^M_g.
\eeq
Until now, we construct the nucleon at quark level with confinement potential and the pion and gluon corrections. Next, we would like to connect such nucleon in nuclear medium with nuclear objects, like nuclear matter and finite nuclei system. A good bridge is the relativistic mean field (RMF) model at hadron level, which is developed based on the one-boson exchange potential between two nucleons. The effective nucleon mass from the quark model will be inserted to the RMF Lagrangian. The nucleon and meson fields will be solved self-consistently, and then, the properties of nuclear many-body system will be obtained. We would like to use a more complicated Lagrangian comparing with MQMC model~\cite{barik13}. The MQMC parameter sets without non-linear terms of $\sigma$ and $\omega$ mesons can provide very good saturation properties of nuclear matter, but for the description of finite nuclei they cannot give consistent results of binding energies and charge radii with the experimental data. For example, with $m_q=300$ MeV in MQMC model, the total energy difference of $^{208}\rm Pb$ between theoretical calculation and experimental data is about $80$ MeV.  Meanwhile, in MQMC model, the effective nucleon mass at saturation density is a little bit larger than the empirical data, which will generate a small spin-orbit splitting comparing with the experimental observation. Therefore, we should introduce the non-linear terms of $\sigma$ and $\omega$ mesons at the nucleon level. At the same time, in this work, we just consider the $\sigma, ~\omega$ and $\rho$ meson exchanging in the QMF Lagrangian ~\cite{shen00}, which is given as
\beq\label{22}
{\cal L}_{\rm QMF} &=&
\bar\psi\bigg[ i\gamma_\mu\partial^\mu-M_N^*
-g_\omega \omega \gamma^0
-g_\rho \rho \tau_3\gamma^0\nn
&&-e\frac{(1-\tau_3)}{2} A \gamma^0
\bigg] \psi \notag \nn
  &&
-\frac{1}{2} (\nabla\sigma)^2
-\frac{1}{2} m_\sigma^2\sigma^2
-\frac{1}{3} g_2\sigma^3
-\frac{1}{4} g_3\sigma^4\nn
&&+\frac{1}{2} (\nabla\omega)^2
+\frac{1}{2} m_\omega^2\omega^2
+\frac{1}{4} c_3\omega^4  \notag \nn
  &	&
+\frac{1}{2} (\nabla\rho)^2
+\frac{1}{2} m_\rho^2\rho^2
+\frac{1}{2}(\bigtriangledown A)^2,
\eeq
where $M^*_N$ is the effective nucleon mass obtained from the quark model, and the coupling constants of $\omega$ and $\rho$ mesons with nucleon can be related with the quark part as, $g_\omega=3g^q_\omega$ and $g_\rho=g^q_\rho$ according to the quark counting rules. $A$ is the electricmagnetic field for the Coulomb interaction between protons. In this Lagrangian, we already consider the static approximation on the mesons so that their time components are neglected. The spatial part of $\omega$ meson disappears for the time reversal symmetry.

From this Lagrangian, the equations of motion of nucleons and mesons will be generated by the Euler-Lagrangian equation,
\beq\label{}
&&\bigg[i\gamma_{\mu}\partial^{\mu}-M_N^*
- g_\omega \omega(r) \gamma^0
-g_\rho \rho(r) \tau_3\gamma^0\nn
&&	-e\frac{(1-\tau_3)}{2} A(r)\gamma^0
 \bigg]\psi=0,\nn
&&\Delta\sigma(r)-m_\sigma^2 \sigma(r)-g_2\sigma^2(r)-g_3 \sigma^3(r)=
\frac {\partial M_N^*}{\partial \sigma}\langle \bar \psi \psi \rangle,\nn
&&\Delta\omega(r)-m_\omega^2 \omega(r)-c_3 \omega^3(r)=
-g_\omega \langle \bar \psi \gamma^0 \psi \rangle,\nn
&&\Delta\rho(r)-m_\rho^2 \rho(r)=
-g_\rho\langle \bar \psi \tau_3\gamma^0 \psi \rangle,\nn
&&\Delta A(r)=
-e\langle \bar \psi \frac{(1-\tau_3)}{2}\gamma^0 \psi \rangle,
\eeq
where, $\frac {\partial M_N^*}{\partial \sigma}$ comes from the quark model and is different from the $g_\sigma$ in RMF model. Here we restrict our consideration to spherically symmetric nuclei and $r$ is the radial coordinate of the nuclear center. These equations of motion can be solved self-consistently in numerical program. From the single particle energies of nucleon, the total energy of whole nucleus can be obtained with the mean field method.

The infinite nuclear matter, which does not really exist in universe, is very helpful for us to understand the basic physics of nuclear many-body system. It has the translational invariance with the infinite system, which removes the partial part on coordinate space. Its Lagrangian density and equations of motion will be written as
\beq\label{28}
\mathcal{L}_{\rm QMF}&=&\bar{\psi}(i\gamma_{\mu}\partial^\mu-M_{N}^{\ast}-g_\omega\omega\gamma^0-g_\rho\rho\tau_3\gamma^0)\psi\nn
                      &&-\frac{1}{2}m^2_\sigma\sigma^2-\frac{1}{3}g_2\sigma^3-\frac{1}{4}g_3\sigma^4\nn
                      &&+\frac{1}{2}m^2_\omega\omega^2
                       +\frac{1}{4}c_3\omega^4+\frac{1}{2}m^2_\rho\rho^2,
\eeq
and
\beq\label{}
&&(i\gamma^{\mu}\del_\mu-M_{N}^{\ast}-g_\omega\omega\gamma^0-g_\rho\rho\tau_3\gamma^0)\psi=0,\nn
&&m_{\sigma}^2\sigma+g_2\sigma^2+g_3\sigma^3=-\frac{\del M_N^*}{\del\sigma}\bra\bar{\psi}\psi\ket,\nn
&&m_{\omega}^2\omega+c_3\omega^3=g_\omega\bra\bar{\psi}\gamma^0\psi\ket,\nn
&&m_{\rho}^2\rho=g_\rho\bra\bar{\psi}\tau_3\gamma^0\psi\ket.
\eeq
From these Lagrangian and equations of motion of nucleon and mesons, the energy density and pressure can be generated by the energy-momentum tensor~\cite{shen02a},
\beq
\mathcal{E}_{\rm QMF}&=&\sum_{i=n,p}\frac{1}{\pi^2}\int^{k^i_F}_0\sqrt{k^2+M^*}k^2dk\nn
&&+\frac{1}{2}m^2_\sigma\sigma^2-\frac{1}{3}g_2\sigma^3+\frac{1}{4}g_3\sigma^4\nn
&&+\frac{1}{2}m^2_\omega\omega^2+\frac{3}{4}c_3\omega^4+\frac{1}{2}m^2_\rho\rho^2,
\eeq
and
\beq
P_{\rm QMF}&=&\frac{1}{3\pi^2}\sum_{i=n,p}\int^{k^i_F}_0\frac{k^4}{\sqrt{k^2+M^*}}dk\nn
&&-\frac{1}{2}m^2_\sigma\sigma^2+\frac{1}{3}g_2\sigma^3-\frac{1}{4}g_3\sigma^4\nn
&&+\frac{1}{2}m^2_\omega\omega^2+\frac{1}{4}c_3\omega^4+\frac{1}{2}m^2_\rho\rho^2.
\eeq

\section{Results and discussions}

The parameters in confinement potential, $(a,V_0)$ are determined by the experiment data of nucleon mass $M_N=939$ MeV and charge radius
${\bra r^2_N\ket}^{1/2}=0.87$ fm in free space. Then we calculate the effective mass $M_N^*$ in nuclear medium as a function of the quark mass correction $\delta m_q$, which is defined as $\delta m_q=m_q-m_q^*=g^q_\sigma\sigma$.
\begin{figure}[H]
	\centering
	\includegraphics[width=9cm]{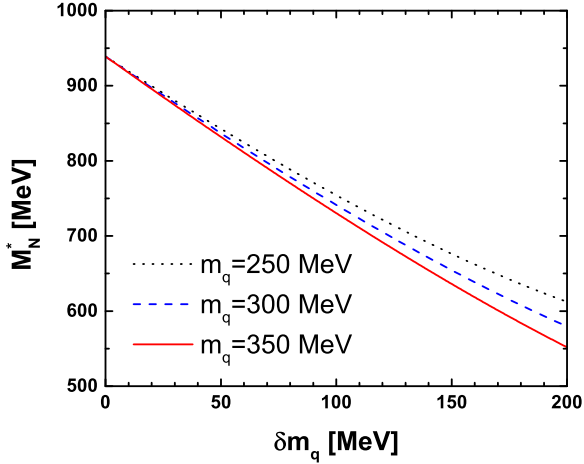}
	\caption{The effective nucleon masses $M^*_N$ as a function of the quark mass correction $\delta m_q$ for quark mass
		$m_q=250$ MeV (dot curve), quark mass $m_q=300$ MeV (dashed curve), and quark mass $m_q=350$ MeV (solid curve), respectively.}
	\label{fig1}
\end{figure}
In Fig.~\ref{fig1}, the effective nucleon masses with different bare quark masses ($m_q=250,~300,$ and $350$ MeV) as functions of quark mass correction are given. The values of $(a,V_0)$ in different quark masses are given in Table \ref{tab1}. At free space ($\delta m_q=0$), their effective masses are completely corresponding the free nucleon mass. With the $\delta m_q$ increasing, the effective nucleon masses will be reduced for the effect of surrounding nucleons. Furthermore, this reduction becomes faster at larger quark mass.

\begin{figure}[H]
	\centering
	\includegraphics[width=9cm]{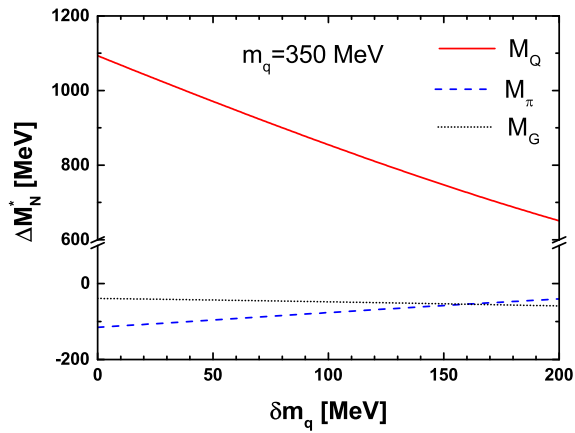}
	\caption{The contributions of confinement potential, pion and gluon corrections in the effective nucleon mass $M^*_N$ as a function of the quark mass correction
		$\delta m_q$ at quark mass $m_q=350$ MeV. $M_Q, ~M_\pi$ and $M_G$ represent the confinement contribution, pion correction and gluon correction, respectively.}
	\label{fig2}
\end{figure}

We also show the contributions from the quark confinement, pion correction and gluon correction on the effective nucleon mass in Fig.\ref{fig2}. The confinement potential provides the positive contribution on effective nucleon  mass and makes the effective mass largely reduced in nuclear medium, while the pion and gluon generate the negative contribution on nucleon mass. The gluon correction is almost not influenced by the nuclear medium, and the pion one increases slowly with $\delta m_q$. Their magnitudes are smaller compared to the one from the confinement potential, which are calculated with the perturbation theory.

After we fixed the parameters of quark confinement potential, the coupling constants between quarks and mesons should be determined by the properties of finite nuclei. In this work, we take the meson masses as $m_\sigma=550$ MeV, $m_\omega=783$ MeV, and $m_\rho=763$ MeV. $g_\sigma^q$, $g_\omega$, $g_\rho$, $g_2$, $g_3$, and $c_3$ will be fitted by the binding energies per nucleon $E/A$ and the charge radii $R_c$ of four closed shell nuclei, \ce{^{40}Ca}, \ce{^{48}Ca}, \ce{^{90}Zr}, and \ce{^{208}Pb} with least squares fitting method. Three parameter sets are achieved at $m_q=250,~300$, and $350$ MeV to study the effect of quark properties on nuclear many-body system. These parameters sets and the corresponding $a$ and $V_0$ in quark confinement potential are listed in Table \ref{tab1}. For the later discussion convenience, we name the first parameter set ($m_q=250$ MeV) in Table \ref{tab1} as QMF-NK1, the second ($m_q=300$ MeV) as  QMF-NK2, and the third ($m_q=350$ MeV) as QMF-NK3.

\begin{table}[H]
	\centering
	\begin{tabular}{l c c c c c c c c}
		\hline
		\hline
		$m_q$&$g_{\sigma}^{q}$&$g_{\omega}$&$g_{\rho}$&$g_2$          &$g_3$      &$c_3$     &a              &$V_0$\\
		(MeV)&                &            &          &$(\rm fm^{-1})$&           &          &$(\rm fm^{-3})$&(MeV)\\
		\hline
		
		250  &5.15871         &11.54726    &3.79601   &-3.52737       &-78.52006  &305.00240 &0.57945       &-24.28660\\
		
		300  &5.09346         &12.30084    &4.04190   &-3.42813       &-57.68387  &249.05654 &0.53430       &-62.25719\\
		
		350  &5.01631         &12.83898    &4.10772   &-3.29969       &-39.87981  &221.68240 &0.49560       &-102.04158\\
		
		\hline
		\hline
	\end{tabular}
	\caption{The parameters in quark and hadron are listed. The first parameter set corresponding to $m_q=250$ MeV is named as QMF-NK1,
		the second $m_q=300$ MeV named QMF-NK2, and the third $m_q=350$ MeV named QMF-NK3.}\label{tab1}
\end{table}

In Table \ref{tab2}, the results of theoretical calculation for the binding energies per nucleon $E/A$ and the charge radii $R_c$ for four spherically symmetric nuclei, \ce{^{40}Ca}, \ce{^{48}Ca}, \ce{^{90}Zr}, and \ce{^{208}Pb} by QMF-NK1, QMF-NK2, and QMF-NK3 are compared with the experimental data. We can find that the results from the QMF-NK3 are closest to the experimental values compared to the other two parameter sets. It demonstrates that the heavier quark mass is more acceptable for the nuclear many-body system.
The calculation without the pion and gluon correction by Shen and Toki \cite{shen00} is also compared and the present results are largely improved. Therefore, it is necessary to include the contributions of pion and gluon to describe the finite nuclei system properly.

{\begin{table}[H]
		\centering
		\begin{tabular}{l c c c c c c c c c c}
			\hline
			\hline
			Model     &~   &            &\multicolumn{2}{c}{$E/A$ (MeV)}&             &~~~&            &\multicolumn{2}{c}{$R_{c}$(fm)}&\\
			&~   &\ce{^{40}Ca}&\ce{^{48}Ca}  &\ce{^{90}Zr}    &\ce{^{208}Pb}&   &\ce{^{40}Ca}&\ce{^{48}Ca}&\ce{^{90}Zr}      &\ce{^{208}Pb}\\
			\hline
			QMF-NK1   &~   &8.62        &8.61          &8.65            &7.92         &   &3.43        &3.47        &4.26              &5.49\\
			
			QMF-NK2   &~   &8.61        &8.61          &8.67            &7.91         &   &3.44        &3.47        &4.26              &5.50\\
			
			QMF-NK3   &~   &8.59        &8.63          &8.68            &7.90         &   &3.44        &3.46        &4.26              &5.50\\
			
   QMF\cite{shen00}   &~   &8.35        &8.43          &8.54            &7.73         &   &3.44        &3.46        &4.27              &5.53\\
			
			Expt.     &~~~~&8.55        &8.67          &8.71            &7.87         &   &3.45        &3.45        &4.26              &5.50\\
			\hline
			\hline
		\end{tabular}
		\caption{The binding energies per nucleon $E/A$ and the rms charge radii $R_c$ with QMF-NK1, QMF-NK2, and QMF-NK3 parameter sets, compared with
			the results in previous QMF model without pion and gluon corrections, and experimental values.}\label{tab2}
	\end{table}}

In Table \ref{tab3}, we also compare the spin-orbit splittings of $^{40}\rm Ca$ and $^{208}\rm Pb$ for QMF-NK1, QMF-NK2, and QMF-NK3 in the present work to the experimental data. With the quark mass increases, the spin-orbit splittings of these nuclei become larger and approaching the experimental data. In RMF model, the spin-splittings are actually strongly dependent on the effective nucleon  mass and have the inverse relation. From Fig.\ref{fig1}, we can observe that the largest quark mass generates the smallest effective nucleon mass at each $\delta m_q$. Therefore, a large quark mass will result in a large spin-orbit splitting.

\begin{table}[H]
	\centering
	\begin{tabular}{l c c c c c}
		\hline
		\hline
		Model       &         &\ce{^{40}Ca}          &                     &\ce{^{208}Pb}        &       \\
		&         &Proton                &Neutron              &Proton               &Neutron\\
		&~~~~     &$(1d_{5/2}-1d_{3/2})$ &$(1d_{5/2}-1d_{3/2})$&$(1g_{9/2}-1g_{7/2})$&$(2f_{7/2}-2f_{5/2})$\\
		\hline
		QMF-NK1     &         &-3.7                  &-3.7                 &-2.4                 &-1.5  \\
		
		QMF-NK2     &         &-4.5                  &-4.5                 &-2.8                 &-1.7  \\
		
		QMF-NK3     &         &-5.1                  &-5.1                 &-3.2                 &-1.9  \\
		
		Expt.       &         &-7.2                  &-6.3                 &-4.0                 &-1.8  \\
		\hline
		\hline
	\end{tabular}
	\caption{The spin-orbit splittings of $^{40}\rm Ca$ and $^{208}\rm Pb$ for QMF-NK1, QMF-NK2, and QMF-NK3, compared with the
		experimental data. All quantities are in MeV.}\label{tab3}
\end{table}

\begin{figure}[H]
	\centering
	\subfigure[]{\label{}\includegraphics[width=8.1cm]{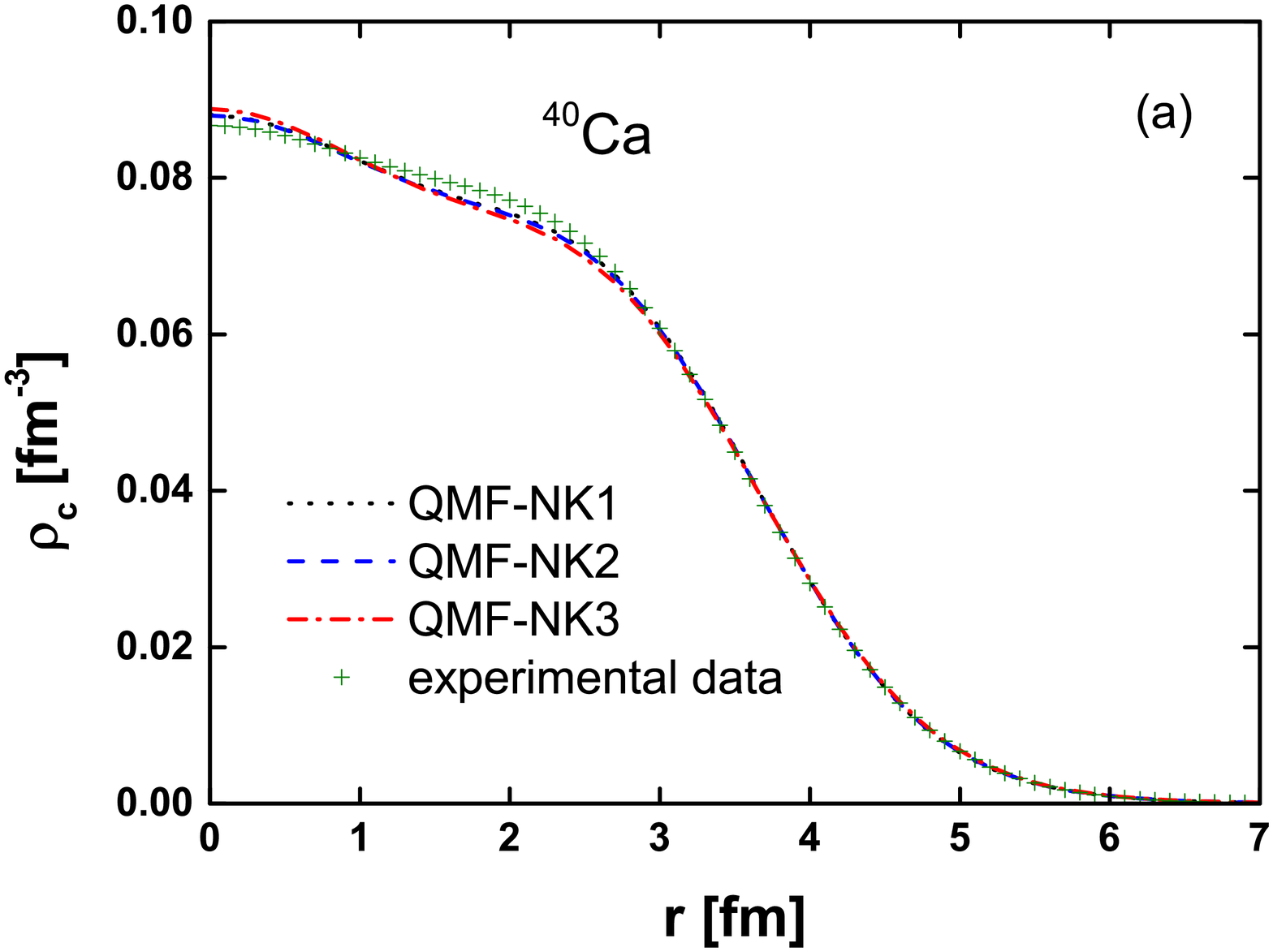}}
	\subfigure[]{\label{}\includegraphics[width=8.1cm]{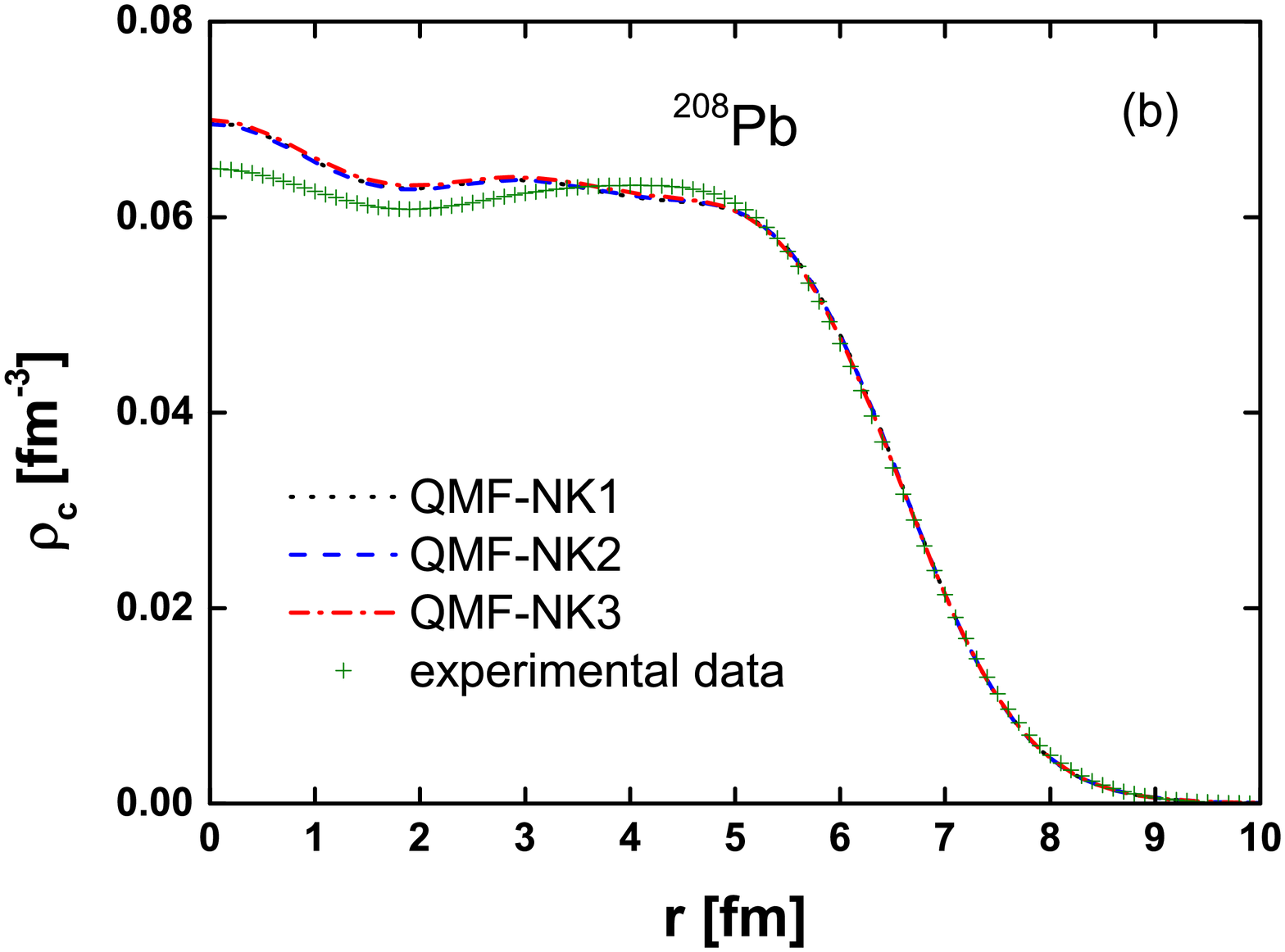}}
	\caption{The charge density distributions of $^{40}\rm Ca$ ((a) panel) and  $^{208}\rm Pb$ ((b) panel) at QMF-NK1, QMF-NK2 and QMF-NK3 compared with the experimental data.}
	\label{fig3}
\end{figure}

\begin{figure}[H]
	\centering
	\includegraphics[width=9cm]{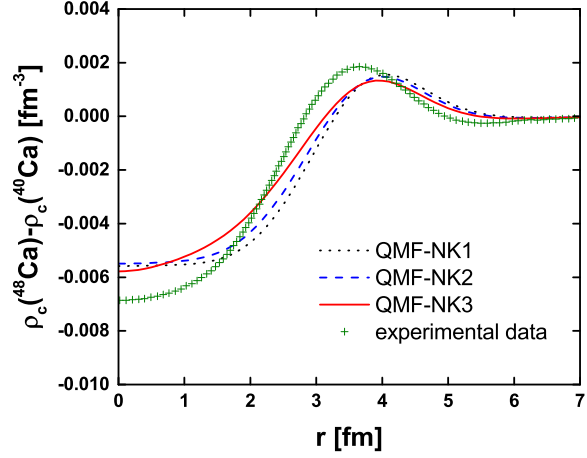}
	\caption{The difference between the $^{48}\rm Ca$ and $^{40}\rm Ca$ charge densities at QMF-NK1, QMF-NK2 and QMF-NK3, compared with the experimental
		data.}
	\label{fig4}
\end{figure}

In Figs. \ref{fig3}, we plot the charge density distributions of $^{40}\rm Ca$ and $^{208}\rm Pb$ for QMF-NK1, QMF-NK2, and QMF-NK3 and compare them with the experimental data. They are almost identical for the three parameter sets and coincident with the behavior of experimental data. In Fig. \ref{fig4}, we also plot the difference between the $^{48}\rm Ca$ and $^{40}\rm Ca$ charge densities for QMF-NK1, QMF-NK2, and QMF-NK3.

Once we fixed the free parameters in QMF Lagrangian from the finite nuclei system, we can apply such parameter sets on the studying of nuclear matter and examine their validity. In Table \ref{tab4}, the various properties of symmetric nuclear matter at saturation density for the three parameter sets are tabulated, such as saturation density, binding energy per particle, incompressibility, symmetry energy and so on. Their detailed expressions can be found in Ref.~\cite{mishra15}. The saturation densities, binding energies and incompressibilities in QMF-NK1, QMF-NK2, and QMF-NK3 are almost the same, which is consistent with the empirical saturation properties of nuclear matter. The symmetry energies and effective masses perform obviously difference for these three parameter sets, which are caused by the coupling constants of $\rho$ meson and quark masses.
\begin{table}[H]
	\centering
	\begin{tabular}{l c c c c c c c c c c c}
		\hline
		\hline
		Model       &~~~&$\rho_0$      &$E/A$&$K_0$           &$J$   &$M^*_N/M_N$&$L^0$&$K^0_{\rm sym}$&$K_{\rm asy}$&$Q_0$ &$K_\tau$\\
		&   &$\rm(fm^{-3})$&(MeV)&(MeV)           &(MeV) &           &(MeV)&(MeV)          &(MeV)        &(MeV) &(MeV)   \\\hline
		
		QMF-NK1     &   &0.154         &-16.3&323             &30.6  &0.70       &84.8 &-28.8          &-537.6       &495.4 &-667.7  \\
		
		QMF-NK2     &   &0.152         &-16.3&328             &32.9  &0.66       &93.7 &-23.5          &-585.7       &221.0 &-648.8  \\
		
		QMF-NK3     &   &0.150         &-16.3&322             &33.6  &0.64       &97.3 &-12.0          &-595.8       &263.0 &-675.3  \\
		
		\hline
		\hline
	\end{tabular}
	\caption{Saturation properties of nuclear matter in the QMF-NK1, QMF-NK2, and QMF-NK3.  }\label{tab4}
\end{table}

\begin{figure}[h]
\centering
  \subfigure[]{\label{}\includegraphics[width=8.1cm]{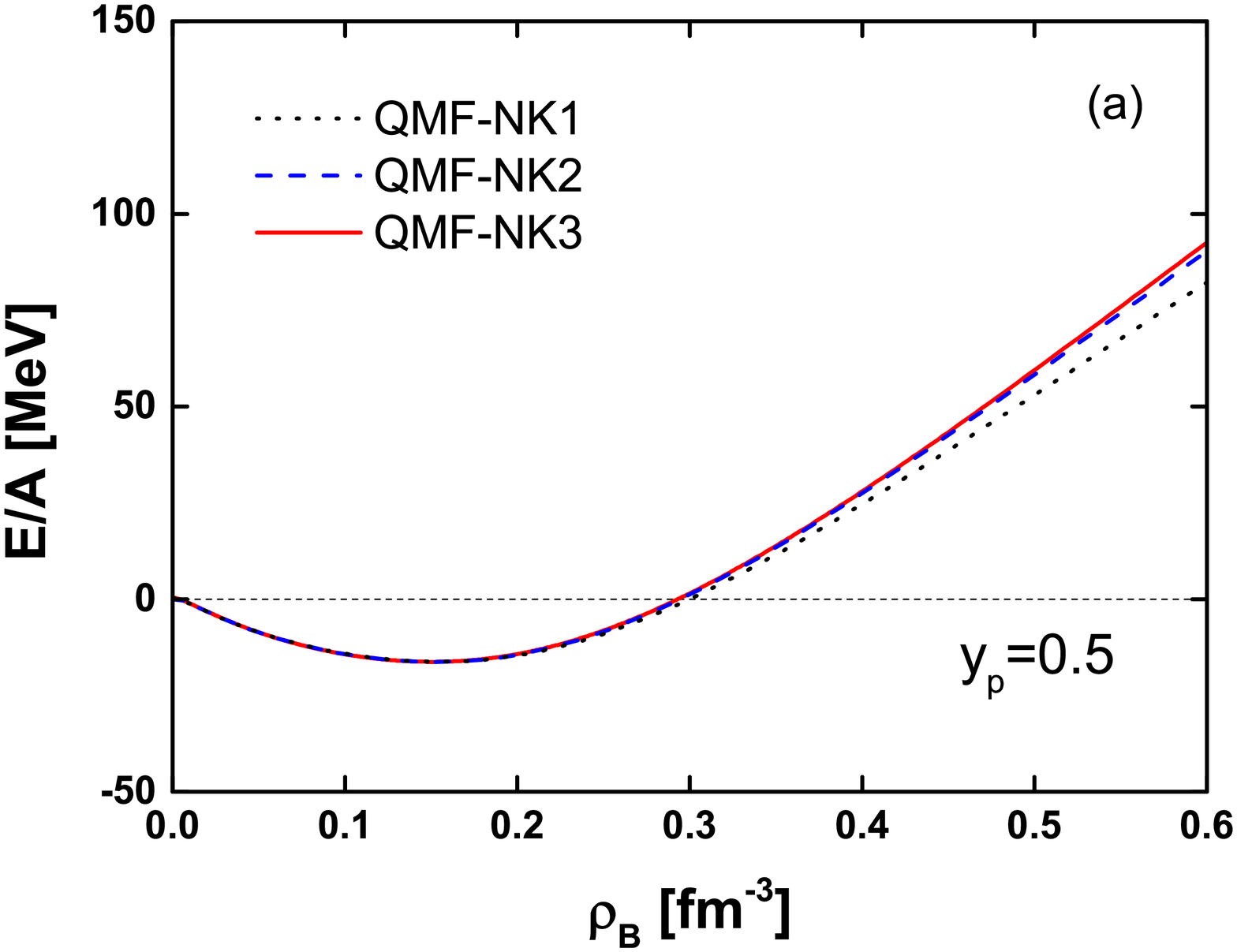}}
  \subfigure[]{\label{}\includegraphics[width=8.1cm]{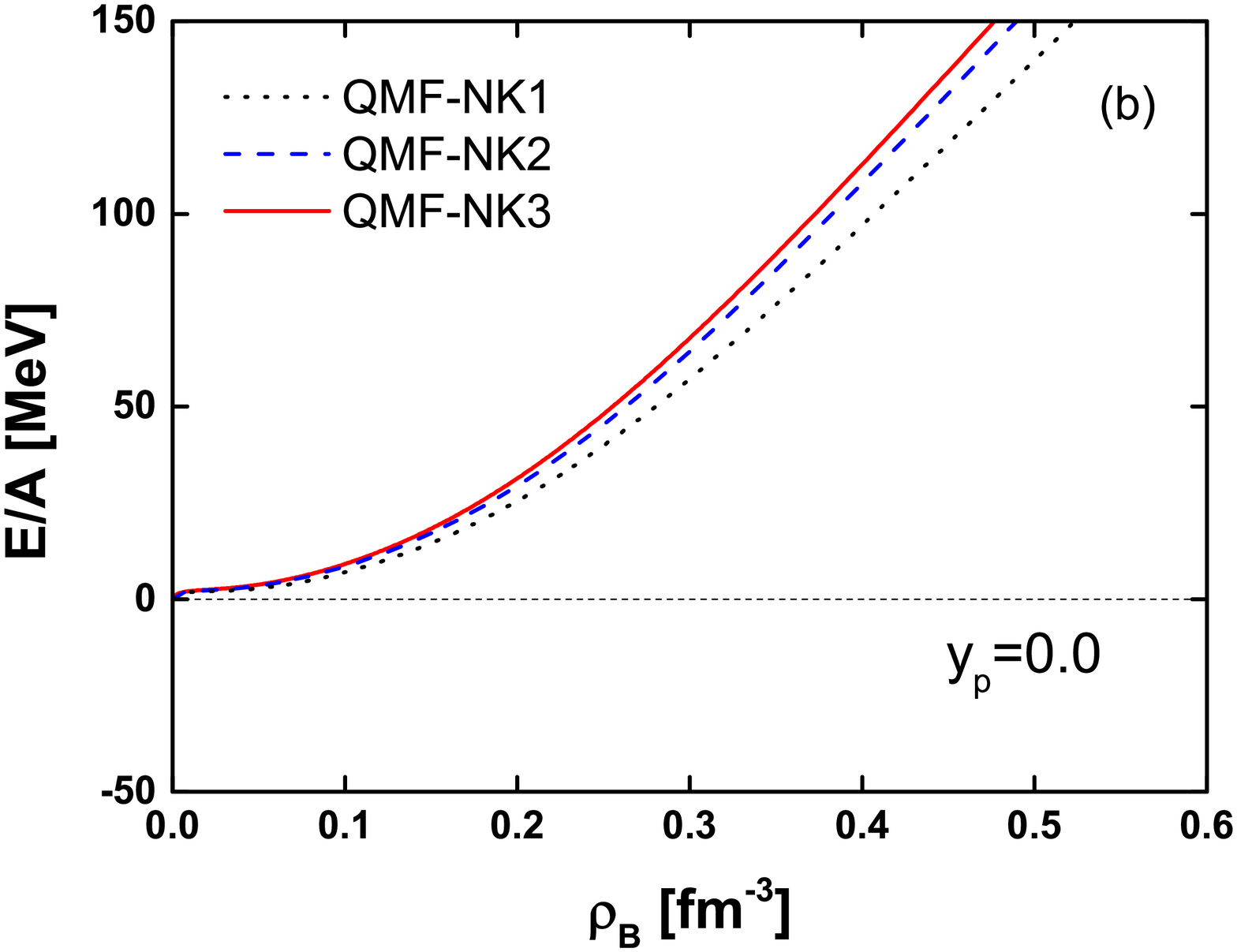}}
\caption{EOSs of symmetric nuclear matter and pure neutron matter in QMF-NK1, QMF-NK2, and QMF-NK3 for (a) $y_p=0.5$ and (b) $y_p=0.0$.}\label{fig5}
\end{figure}

In Fig. \ref{fig5}, we plot nuclear matter binding energy as a function of density in the three parameter sets for symmetric nuclear matter and pure neutron matter. At low densities, the equations of state (EOSs) for different sets are identical. With the density increasing, the EOS becomes softer for lower quark mass.

\begin{figure}[H]
	\centering
	\includegraphics[width=9cm]{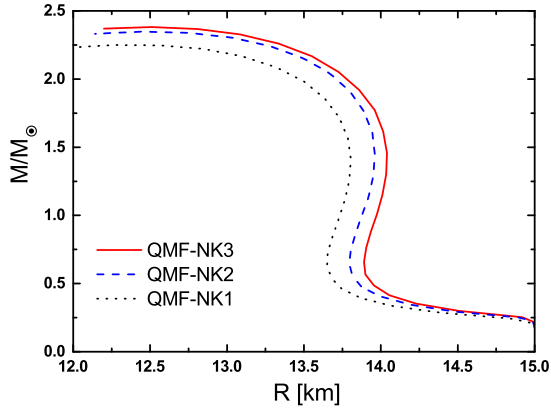}
	\caption{The mass-radius relations for neutron stars in QMF-NK1, QMF-NK2, and QMF-NK3.}
	\label{fig6}
\end{figure}

The neutron star as a natural laboratory is a very good object to check the nuclear theoretical model. We would like to calculate the properties of neutron star with QMF model and show the mass-radius relations for the neutron stars in Fig. \ref{fig6}. The maximum masses of neutron star in this work are between $2.25M_\odot$ and  $2.38M_\odot$, which satisfy the constraint of present astronomical observation data about $2M_\odot$~\cite{demorest10}. However, the previous QMF model without pion and gluon corrections, could not provide the large neutron star mass~\cite{hu14b}.

\section{Conclusion}

We studied the properties of finite nuclei and infinite nuclear matter in terms of quark mean field (QMF) model with the effects of pion and gluon. In the QMF model, the nucleon is made up of three constituent quarks with a confinement potential. Due to the chiral symmetry in QCD theory and quark exchange interaction, the corrections of pion and gluon have been considered in a perturbation manner to obtain the effective nucleon  mass in the medium. The strength of confinement potential is determined by the mass and charge radius of free nucleon. For hadron part in QMF model, the nucleon is combined through the meson exchanging between quarks. Comparing with the previous MQMC model~\cite{barik13}, we also included the non-linear terms of $\sigma$ and $\omega$ mesons. Their coupling constants ($g_\sigma^q$, $g_\omega$, $g_\rho$, $g_2$, $g_3$, and $c_3$) were determined by fitting the experimental data of the binding energies per nucleon $E/A$ and the charge radii $R_c$ of four closed shell nuclei, $^{40}\rm Ca$, $^{48}\rm Ca$, $^{90}\rm Zr$, and $^{208}\rm Pb$. Finally, we obtained three parameter sets with different quark confinement potentials named as QMF-NK1, QMF-NK2, and QMF-NK3. The present QMF model is largely improved to describe the properties of finite nuclei compared to previous QMF version without pion and gluon corrections and MQMC model. The MQMC parameter sets without non-linear terms of $\sigma$ and $\omega$ mesons could provide very good saturation properties of nuclear matter, but for the description of finite nuclei they cannot give consistent results of binding energies and charge radii with the experimental data. For example, with $m_q=300$ MeV in MQMC model, the total energy difference of $^{208}\rm Pb$ between theoretical calculation and experimental data is about 80 MeV. After we introduced the non-linear terms of $\sigma$ and $\omega$ mesons, this difference was reduced to 6 MeV with QMF-NK3 parameter set, while the saturation properties of nuclear matter were still kept very well. Furthermore, we also calculated the spin-orbit splittings for $^{40}\rm Ca$ and $^{208}\rm Pb$, and the charge density distributions in comparison with the experimental data. The spin-orbit splittings in our work were largely improved comparing with the ones in MQMC model, where effective nucleon mass was very large at saturation density to generate a small spin-orbit splittings of finite nuclei. We also found that the spin-orbit splittings increased with the quark mass, since a smaller effective nucleon mass usually generates strong spin-orbit force in RMF framework.

We also applied these parameter sets on the study of infinite nuclear matter. The various saturation properties of symmetric nuclear matter are consistent with the empirical data. The obvious difference in QMF-NK1, QMF-NK2, and QMF-NK3 reflected on the symmetry energies and effective nucleon masses at saturation density, which were strongly dependent on the strength of $\rho$ meson coupling with quarks and the quark mass. The equations of state (EOSs) of symmetry nuclear matter and pure neutron matter were given, where the EOS with large quark mass became stiffer. The mass-radius relations of neutron stars were calculated. The maximum neutron star masses in the present QMF model were around $2.25M_\odot$ to $2.38M_\odot$, which satisfied the recent constraint of astrophysics observation. Recently, there are many discussions on the hyperon degrees of freedom for the massive neutron stars, which is called as "hyperon puzzle". In future, we will include the strange quark degree of freedom in quark level to generate more baryon states, like $\Lambda,~\Sigma$ and $\Xi$ to study the hyperon degrees of freedom in neutron star.

With the pion and gluon corrections, the QMF model could treat the finite nuclei and nuclear matter better. Although it is a long road to describe the nuclear many-body system with QCD theory directly, the influence of quark level on nuclear structure was found out in the present QMF model, such as effective nucleon mass. With the energy and density increasing, the strangeness degree of freedom will appear in nuclear physics. We will consider more baryon states in such model and study their roles in hypernuclei and neutron stars.

\section*{Acknowledgments}
We are grateful to the detailed fruitful discussions with Prof. Hiroshi Toki and Dr. Ang Li. This work was supported in part by the National Natural Science Foundation of China (Grant No. 11375089 and Grant No. 11405090).

\end{document}